\documentclass[structabstract]{aa}  
\usepackage{graphicx}
\usepackage{txfonts}
\usepackage{rotating}
\usepackage{natbib,twoopt}
\usepackage{breqn}

\usepackage{subfig}

\begin{document}

% Lucky Imaging observations of transit host stars to find close stellar sources
% A Lucky Imaging search for stellar companions to transiting planet host stars II: new observations
\title{A Lucky Imaging search for stellar sources near $74$ transit hosts\thanks{ Based on observations collected at the German-Spanish Astronomical Center, Calar Alto, jointly operated by the Max-Planck-Institut f\"ur Astronomie Heidelberg and the Instituto de Astrof\'{\i}sica de Andaluc\'{\i}a (CSIC).}}

   \author{ Maria W\"ollert\inst{1} \and
            Wolfgang Brandner\inst{1} 
           }

   \institute{ Max-Planck-Institut f\"ur Astronomie, K\"onigstuhl 17,
               69117 Heidelberg, Germany\\
               \email{woellert@mpia.de} 
              }

   \date{Received ---; accepted ---}

% \abstract{}{}{}{}{} 
% 5 {} token are mandatory
  
     \abstract
   {Many transiting planet host stars lack high resolution imaging and thus close
stellar sources can be missed. Those unknown stars potentially bias the
derivation of the planetary and stellar parameters from the transit light curve, no
matter if they are bound or not. In addition, bound stellar companions interact
gravitationally with the exoplanet host star, the disk and the planets and can thus
influence the formation and evolution of the planetary system strongly.}
   {We extended our high-resolution Lucky Imaging survey for close stellar sources by $74$ transiting planet
host stars. $39$ of these stars lack previous high-resolution imaging, $23$ are follow up observations of companions or companion candidates, and the remaining stars have been observed by others with AO imaging though in different bands. We determine the separation of all new and known companion candidates and estimate the flux ratio in the observed bands.
}
   {All observations were carried out with the Lucky Imaging camera AstraLux Norte at the Calar Alto $2.2\,$m telescope in $i^{\prime}$ and $z^{\prime}$ passbands. }
   {We find new stellar sources within $1''$ to HAT-P-27, HAT-P-28, HAT-P-35, WASP-76, and WASP-103 and between $1''$ and $4 \arcsec$ to HAT-P-29, and WASP-56. }
   {}

   \keywords{Techniques: high angular resolution --
             Binaries: visual --
             Planetary systems
               }
   \titlerunning{A LI search for stellar sources near TEP hosts}
   \maketitle
%
%________________________________________________________________

\section{Introduction}

During the last $15\,$years, more than $1000$ confirmed and several 1000 candidate exoplanets have been found by ground- and space-based transit searches as HATNet \citep{Bakos04}, SuperWasp \citep{Pollacco06}, CoRoT \citep{Baglin06}, and Kepler \citep{Borucki10,Batalha13,Burke14}. Transiting exoplanets (TEPs) offer the unique opportunity to determine a variety of planetary properties as true mass, mean density and surface gravity. They also allow to characterize the planet's atmosphere through spectroscopy, to determine the planet's temperature in secondary eclipse observations, and to measure the angle between the orbital plane and the stellar rotation axis via the Rossiter-McLaughlin effect \citep{Winn05}.

As many transiting planet follow-up observations were limited in angular resolution either due to instrumental limits (like, e.g., SPITZER/IRAC) or - in case of ground-based follow-up - seeing limited, care has to be taken about missing a blended close star. This is especially true for faint sources as bright stars may be recognized in follow-up spectra. Unknown, close stars add a constant flux to the light-curve which bias both, primary and secondary eclipse measurements. In the first case, the additional source leads to an underestimate of the planetary radius and consequently an overestimate of the planetary density. In the second case, the planet infrared emission spectrum can be underestimated by several tens of percent \citep[e.g][]{Crossfield12}.

Finding close stellar sources to transiting exoplanet host stars is, however, not only crucial to determine the planetary parameters correctly, but also to understand the influence of binarity on the formation and evolution of planetary systems. Even though not all of the detected, close stars are gravitationally bound, a lot of them are as has been shown via multi-epoch high-resolution observations \cite[e.g][]{Narita12, Bergfors13, NGO15}. The effects of binarity may be manifold: Stellar companions might stir \citep{Mayer05}, tilt \citep{Batygin12} or truncate the protoplanetary disk \citep{Artymowicz94} or they can interact with the formed planets via, e.g., the Lidov-Kozai mechanism or other secular interactions \citep{Wu03,Fabrycky07,Naoz11}. Stellar companions may be thus one important cause for the observed variety of planetary system architectures. 

Several groups have already done systematic surveys for stellar companions using either the Lucky Imaging method \citep[e.g.][]{Daemgen09, Faedi13, Bergfors13, Lillo-Box12, Lillo-Box14, Woellert15}, speckle imaging \citep[e.g.][]{Howell11,Horch14,Kane14,Everett15}, AO-assisted imaging on its own, or combined with radial velocity methods \citep[e.g.][]{Adams12, Adams13, Guenther13, Dressing14, Law14, Wang14, NGO15}, or search for colour-dependency of the transit depths \citep[e.g.][]{Colon12,Desert15}. However, more and more transiting exoplanets are found and their precise characterisation will enable us to get a more precise view at the important mechanisms that shape planetary systems.

In this paper we present the results of our ongoing effort to find stellar sources close to TEP host stars. The observations and data reductions were performed similarly to our previous paper \cite{Woellert15} and are briefly described in Section $2$. In Section $3$ we present the astrometric and photometric properties of the observed sources and we summarize our findings in Section $4$.

\section{Observations and data reduction}

\subsection{Sample selection}

The initial motivation for our survey was to focus on TEPs with already existing measurements of the Rossiter-McLaughlin effect, and to explore a possible relationship of the angle defined by the spin vectors of the TEP host star and the planetary orbit, and the presence or absence of a stellar companion. This selection criterion was later relaxed to include all TEP host stars sufficiently bright (i $\le$ 13\,mag) to facilitate high-quality Lucky Imaging. The majority of the targets were selected from TEPs either identified by the SuperWASP or the HatNet project. This was complemented by TEP hosts identified in other ground- or space-based surveys. We focused on stars without previous high-angular resolution observations, as well as on TEP host stars with previous detections of stellar companion candidates in order to derive constraints on the relative astrometry between TEP host and stellar companion candidate.

\subsection{Lucky Imaging with AstraLux at Calar Alto}

All observations were carried out at Calar Alto with the $2.2\,$m telescope in combination with the Lucky Imaging camera AstraLux Norte \citep{Hormuth08} during two observing runs, one night in October 2014 and three nights in March 2015. The targets were observed in SDSS $i^{\prime}$ and $z^{\prime}$-passbands using the same set-up as described in \cite{Woellert15}. The field of view was $12''$ by $12''$ with the exoplanet host star separated at least $4''$ from the image borders. Depending on the target brightness and observing conditions we took between $10000$ and $54000$ individual frames with exposure times of $15\,$ms each so that the probability of getting a stable speckle pattern is sufficiently large. The individual AstraLux images are dark subtracted and flat-fielded. For the data analysis we chose the $10\%$ of images with the highest Strehl ratio and combined them using the shift-and-add technique. 

In order to precisely measure the separation and position angle of the companion candidates we took at least $3$ images of the globular cluster M$\,13$ each night. Using IRAF imexamine we determined the detector position of $5$ widely separated stars in the field and calculated their separation and rotation angle pairwise. The result was compared to the values from high-quality astrometric HST/ACS observations. As the instrument was not unmounted during the observing nights in March 2015, we used the images of all three nights for the calibration of that run. The plate scale and detector rotation were $23.46 \pm 0.01$ mas/px and $1.7^{\circ} \pm 0.1^{\circ}$ west of north and $23.51 \pm 0.01$ mas/px and $2.0^{\circ} \pm 0.1^{\circ}$ west of north for the observations in October 2014 and March 2015, respectively.

\subsection{Photometry and astrometry}

To find all stellar sources, also those which are faint and close to the transiting planet host star, we first subtracted the point spread function (PSF) of the TEP host. As the PSFs vary significantly from image to image and no standard PSF can be used for all observations, we fitted a theoretical model PSF to each star. The theoretical model PSF comprises the PSF from an ideal telescope without atmosphere which is then convolved with a Gaussian and finally added to a Moffat profile. The model also weights the two PSF components \citep[see][for more information on the procedure]{Woellert14}. If a companion candidate was found, we fitted a scaled and differently weighted PSF to it to determine its position and the flux ratio of the two components. We used the PSF subtracted images additionally to calculate the $5 \sigma$-contrast curve. For this purpose, we divided the flux in a box of $5 \times 5\,$pixel around each pixel by the flux of star in a similarly sized box centred on the peak in the original, not PSF subtracted image. The contrast at a specific separation is then calculated as the median of the contrast of all pixels at the corresponding separation. The contrast at $r=0.25'', 0.5'', 1.0''$ , and $2.0''$ of targets with candidate companions is given in Table~\ref{with_comp}, the contrast for all other targets is given in Table~\ref{no_comp}. Outside of $2.0''$, the contrast decreases only very little and the value given for $2.0''$ can be assumed.  

The flux ratio in both passbands was measured using the IDL routine APER which performs aperture photometry. We used an aperture size of $4.5\,$px which is about the full width half maximum of the stellar PSFs. In contrast to the usual approach, the sky background was not measured in a close annulus around the star, but in a $50 \times 50\,$px sized box in one corner of the image without stellar source for the primary and at the opposite position of the TEP host with the same distance and aperture size for the fainter companion. This ensures that the flux contribution of the brighter TEP host is accurately subtracted from those of the fainter source as our PSFs are almost point symmetrical in shape. The uncertainties of the flux ratios are propagated from the statistical photometric errors given by APER and systematic errors from using this method. The latter were estimated by comparing the results obtained by using different aperture sizes as well as the flux ratios determined by PSF-fitting.

\section{Results}

\begin{figure*}[htb]
    \begin{center}
        \includegraphics[width=1.0\textwidth]{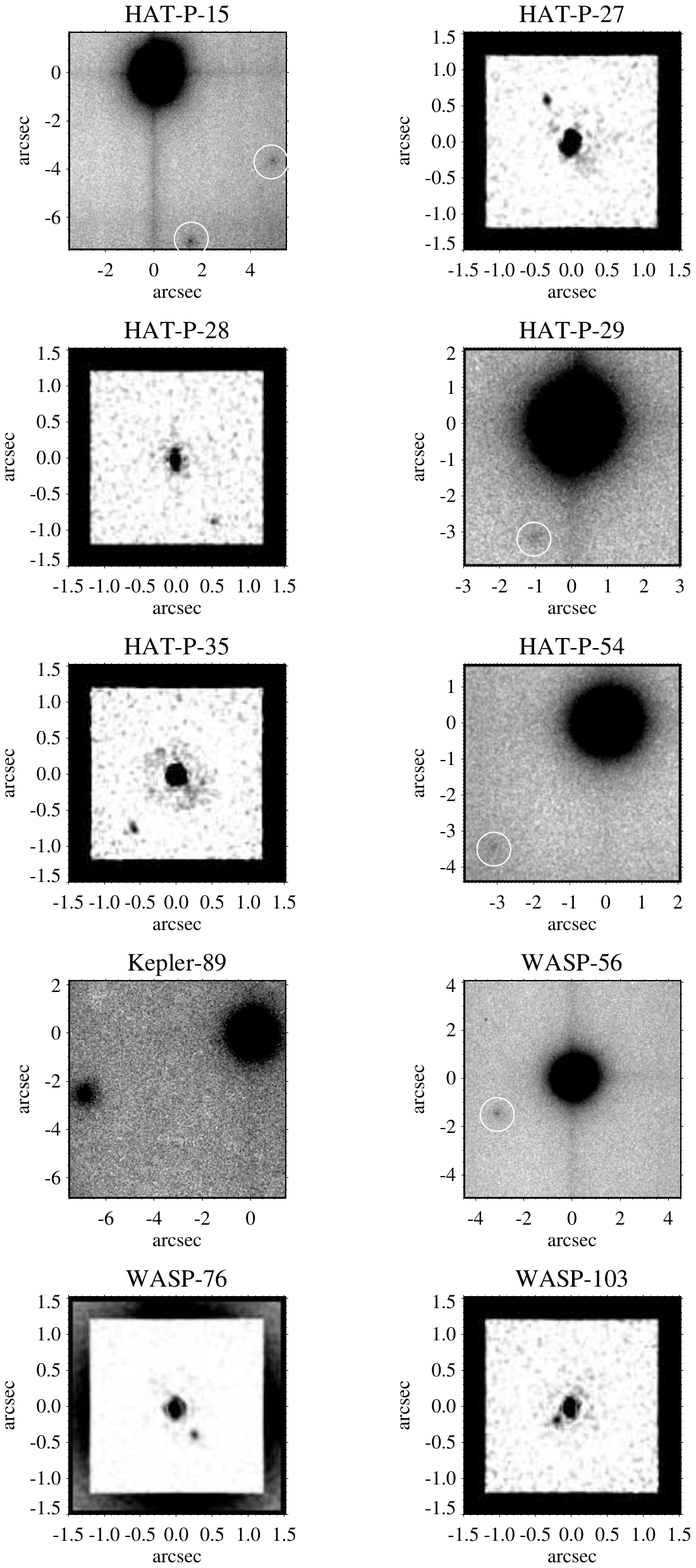}
    \end{center}
    \caption[]{The $z^{\prime}$ filter images of the $10$ exoplanet host stars for which new companion candidates have been detected with exception of HAT-P-35 for which the $i^{\prime}$-image is shown. The grey scale is proportional to the square root of the count. To improve the visibility of the close companions to HAT-P-27, HAT-P-28, HAT-P-35, WASP-76, and WASP-103, we additionally applied unsharp masking. The orientation is identical for all images with North up and East to the left.} 
\label{fig1}
\end{figure*}

To the $74$ observed TEP host stars we find new companion candidates to HAT-P-27, HAT-P-28, HAT-P-35, WASP-76, and WASP-103 within $1''$, to HAT-P-29 and WASP-56 within $4''$, and the candidates of  HAT-P-15, HAT-P-54, and Kepler-89 are situated outside of $4''$. Images in $z^{\prime}$ of all these sources can be found in Figure~\ref{fig1}. In addition, we did follow-up observations of $23$ already known companion candidates. The astrometric positions and flux ratios in $i^{\prime}$ and $z^{\prime}$ of all companion candidates can be found in Table~\ref{results}.

As can be seen in Table~\ref{results}, most sources appear to be redder than the primary which would be expected for a lower mass companion. The uncertainties are, however, too large to allow a precise estimate of the spectral type. To achieve this, adaptive optics based observations or spectra would be needed. The knowledge of the companion candidate spectral type would then allow to correct the planetary parameter and infrared emission spectra of the new close companions, as well as to compare their photometric distance to those of the TEP host star to investigate whether the sources may be gravitationally bound or not. For this purpose their astrometric position needs to be followed up in the upcoming years as well.

\begin{table*}[htb]
	\caption[]{TEP hosts with detection, radial contrast limits, references to other high-resolution imaging papers if available}
    \label{with_comp}
    \begin{center}
	\begin{tabular}{l l l l l l}
        \hline
        \hline
         		&  \multicolumn{4}{c}{$5 \sigma$ detection limit ($\Delta z^{\prime}$ [mag])} &   \\
       	Name 	& $0.25''$      & $0.5''$        & $1.0''$      & $2.0''$ &  ref. \\
		\hline
        \multicolumn{5}{l}{New companion candidates} \\
        \hline
		HAT-P-15	&	4.28	&	5.37	&	6.40	&	6.90	&	\cite{NGO15}$^\ast$  \\
		HAT-P-27/WASP-40	&	3.90	&	4.82	&	5.75	&	6.20	&	\cite{Woellert15}$^\star$ \\
		HAT-P-28	&	3.24	&	3.94	&	4.76	&	5.12	&	\\
		HAT-P-29	&	4.05	&	4.72	&	5.45	&	5.95	&	\cite{NGO15}$^\star$\\
		HAT-P-35	&	2.81	&	3.37	&	3.94	&	4.11	&	\\
		HAT-P-54	&	3.57	&	4.41	&	5.13	&	5.52	&	\\
		Kepler-89	&	3.10	&	3.60	&	4.28	&	4.69	&	\cite{Adams12}$^\ast$; \cite{Lillo-Box14}$^\ast$\\
		WASP-56	&	4.08	&	4.89	&	5.78	&	6.11	&	\\
		WASP-76	&	3.81	&	4.79	&	6.21	&	7.18	&	\\
		WASP-103	&	3.75	&	4.55	&	5.48	&	5.90	&	\\
		\hline
        \multicolumn{5}{l}{Known companion candidates} \\
        \hline
		CoRoT-02	&	2.97	&	3.46	&	4.16	&	4.78	&	\cite{Alonso08, Faedi13, Woellert15}\\
		CoRoT-03	&	2.85	&	3.22	&	3.65	&	3.97	&	\cite{Deleuil08, Faedi13, Woellert15}\\
		CoRoT-11	&	3.25	&	3.85	&	4.65	&	4.97	&	\cite{Gandolfi10, Woellert15}\\
		HAT-P-20	&	3.64	&	4.30	&	5.22	&	5.97	&	\cite{NGO15}$^\ast$; \cite{Bakos11}$^\dagger$ \\
		HAT-P-24	&	3.51	&	4.39	&	5.11	&	5.67	&	\cite{NGO15} \\
		HAT-P-30	&	4.01	&	5.06	&	6.09	&	6.72	&	\cite{Adams13, NGO15}\\
		HAT-P-32	&	3.78	&	4.61	&	5.35	&	5.84	&	\cite{Adams13, NGO15}; \cite{Woellert15}$^\star$\\
		HAT-P-41	&	3.24	&	3.62	&	4.37	&	4.97	&	\cite{Hartman12, Woellert15}\\
		KELT-2	&	3.75	&	4.61	&	6.13	&	7.03	&	\cite{Beatty12}$^\dagger$\\
		KELT-3	&	3.95	&	4.75	&	5.45	&	5.95	&	\cite{Pepper13}\\
		Kepler-13	&	3.33	&	4.11	&	4.87	&	5.54	&	\cite{Santerne12}\\
		KIC10905746	&	3.33	&	3.78	&	4.56	&	5.09	&	\cite{Fischer12}\\
		LHS-6343	&	2.72	&	3.19	&	3.73	&	4.40	&	\cite{Johnson11, Montet15}\\
		TrES-4	&	6.13	&	6.09	&	6.07	&	6.13	&	\cite{Daemgen09, Bergfors13, Faedi13} ;\\ & & & & & \cite{Woellert15, NGO15}\\
		WASP-11	&	3.80	&	4.81	&	5.70	&	6.25	&	\cite{NGO15}\\
		WASP-12	&	4.03	&	4.78	&	5.56	&	6.12	&	\cite{Crossfield12, Bergfors13, Bechter14}\\
		WASP-14	&	4.26	&	5.29	&	6.82	&	7.70	&	\cite{Woellert15, NGO15}\\
		WASP-33	&	3.33	&	4.77	&	6.64	&	7.90	&	\cite{Moya11, Adams13}\\
		WASP-36	&	3.38	&	4.34	&	4.84	&	5.16	&	\cite{Smith12}$^\dagger$ \\
		WASP-70	&	3.07	&	3.52	&	4.26	&	4.90	&	\cite{Anderson14}$^\dagger$\\
		WASP-77	&	3.71	&	4.54	&	5.60	&	6.42	&	\cite{Maxted13}\\
		WASP-85	&	4.25	&	5.26	&	6.21	&	7.12	&	\cite{Brown14}$^\dagger$ \\
		XO-3	&	4.11	&	4.91	&	5.79	&	6.33	&	\cite{Bergfors13}; \cite{Adams13, NGO15}$^\ast$ \\
		\hline
	\end{tabular}
	\begin{quote}	
		$^\ast$: Outside FoV, $^\star$: no detection, $^\dagger$: seeing limited observation or catalog data
	\end{quote}
	\end{center}
\end{table*}

\begin{table*}[htb]
	\caption[]{TEP hosts with no detection, radial contrast limits, references to other high-resolution imaging papers if available}
    \label{no_comp}
    \begin{center}
	\begin{tabular}{l l l l l l}
        \hline
        \hline
         		&  \multicolumn{4}{c}{$5 \sigma$ detection limit ($\Delta z^{\prime}$ [mag])} &  \\
       	Name 	& $0.25''$      & $0.5''$        & $1.0''$      & $2.0''$ & ref. \\
		\hline
		55_Cnc & 2.90 & 2.70 & 3.04 & 3.85 &\cite{Roell12} \\
		CoRoT-01 & 3.23 & 3.81 & 4.34 & 4.70 &\cite{Adams13} \\
		CoRoT-07 & 3.81 & 4.47 & 5.26 & 5.80 &\cite{Guenther13} \\
		CoRoT-24 & 2.78 & 3.26 & 3.44 & 3.51 &\cite{Guenther13} \\
		EPIC-201367065 & 4.16 & 4.98 & 5.99 & 6.69 &\cite{Crossfield15} \\
		EPIC-201505350 & 3.77 & 4.56 & 5.43 & 5.77 & \\
		GJ3470 & 4.01 & 4.97 & 5.61 & 5.96 & \\
		HAT-P-09 & 3.31 & 4.24 & 4.77 & 5.24 & \\
		HAT-P-25 & 3.80 & 4.72 & 5.57 & 5.80 &\cite{Adams13} \\
		HAT-P-33 & 3.86 & 4.60 & 5.29 & 5.80 &\cite{Adams13, NGO15} \\
		HAT-P-38 & 3.59 & 4.46 & 5.22 & 5.61 & \\
		HAT-P-39 & 3.28 & 4.23 & 4.87 & 5.32 & \\
		HAT-P-42 & 1.79 & 2.25 & 2.95 & 3.44 & \\
		HAT-P-43 & 3.33 & 4.18 & 4.82 & 5.12 & \\
		HAT-P-44 & 3.73 & 4.64 & 5.40 & 5.67 & \\
		HAT-P-45 & 3.43 & 4.12 & 4.75 & 5.11 & \\
		HAT-P-46 & 3.97 & 4.64 & 5.50 & 5.94 & \\
		HAT-P-49 & 2.83 & 3.38 & 3.98 & 4.64 & \\
		KELT-1 & 3.65 & 4.19 & 5.14 & 5.88 &\cite{Siverd12} \\
		KELT-6 & 4.07 & 5.14 & 6.37 & 7.19 &\cite{Collins14} \\
		KELT-7 & 3.80 & 5.45 & 6.78 & 7.70 &\cite{Bieryla15} \\
		Kepler-63 & 3.59 & 4.36 & 5.36 & 5.73 &\cite{Sanchis-Ojeda13} \\
		KOI-1474 & 2.79 & 3.33 & 3.79 & 4.16 & \\
		Qatar-2 & 3.59 & 4.32 & 5.12 & 5.47 & \\
		TrES-5 & 2.94 & 3.48 & 3.93 & 4.30 & \\
		WASP-30 & 3.28 & 3.72 & 4.63 & 5.11 & \\
		WASP-32 & 3.47 & 4.10 & 5.11 & 5.56 & \\
		WASP-35 & 3.72 & 4.42 & 5.37 & 5.98 & \\
		WASP-43 & 3.80 & 4.63 & 5.37 & 5.96 & \\
		WASP-44 & 2.85 & 3.37 & 4.08 & 4.47 & \\
		WASP-50  & 3.42 & 3.97 & 4.79 & 5.25 & \\
		WASP-54 & 4.04 & 5.02 & 6.22 & 6.93 & \\
		WASP-57 & 3.56 & 4.63 & 5.29 & 5.46 & \\
		WASP-65 & 3.89 & 4.73 & 5.63 & 6.09 & \\
		WASP-69 & 3.62 & 4.39 & 5.13 & 5.65 & \\
		WASP-71 & 3.59 & 4.17 & 5.22 & 5.94 & \\
		WASP-82 & 3.89 & 4.80 & 5.81 & 6.64 & \\
		WASP-84 & 3.90 & 4.91 & 5.81 & 6.66 & \\
		WASP-90  & 2.83 & 3.33 & 3.85 & 4.36 & \\
		WASP-104 & 3.99 & 4.88 & 5.75 & 6.31 & \\
		WASP-106 & 3.77 & 4.56 & 5.41 & 5.96 & \\
		\hline
	\end{tabular}
	\end{center}
\end{table*}

 \begin{table*}[htb]
 	\caption[]{TEP hosts with candidate companions, observing date, inferred astrometric position and flux ratio in the observed passbands. In the last column we indicate whether the companion was announced previously.}
     \label{results}
     \begin{center}
 	 \begin{tabular}{l l l l l l l}
         \hline
         \hline
		Name		&	Date of obs.	& 	Sep [$\arcsec$]		  &	PA [$^\circ$]		  & 	$\Delta \, i^{\prime}$ [mag] & 	$\Delta \, z^{\prime}$	[mag] & new?		\\
		\hline
		CoRoT-02	&	21.10.2014	& $	4.109	\pm	0.025	$ & $	208.56	\pm	0.14	$ & $	3.35	\pm	0.15	$ & $	3.07	\pm	0.15	$ & \\
		CoRoT-03	&	21.10.2014	& $	5.221	\pm	0.013	$ & $	175.62	\pm	0.55	$ & $	3.39	\pm	0.25	$ & $	3.48	\pm	0.36	$ & \\
		CoRoT-11	&	21.10.2014	& $	2.540	\pm	0.005	$ & $	307.38	\pm	0.17	$ & $	2.27	\pm	0.09	$ & $	2.14	\pm	0.09	$ & \\
		HAT-P-15 south west	&	21.10.2014	& $	6.253	\pm	0.026	$ & $	233.42	\pm	1.54	$ & $	7.17	\pm	0.49	$ & $	6.74	\pm	0.45	$ & yes \\
					&	07.03.2015	& $	6.136	\pm	0.014	$ & $	235.71	\pm	1.02	$ & $	-					$ & $	6.79	\pm	0.22	$ & \\
		HAT-P-15 south	&	07.03.2015	& $	7.091	\pm	0.014	$ & $	194.23	\pm	1.02	$ & $	-					$ & $	6.66	\pm	0.20		$ & yes \\
		HAT-P-20	&	21.10.2014	& $	6.925	\pm	0.012	$ & $	321.10	\pm	0.11	$ & $	2.01	\pm	0.08	$ & $	1.67	\pm	0.08	$ & \\
		HAT-P-24	&	06.03.2015	& $	4.965	\pm	0.008	$ & $	171.32	\pm	0.60	$ & $	6.01		\pm	0.18	$ & $	5.80		\pm	0.24	$ & \\
		HAT-P-27/WASP-40	&	27.06.2013 $^\dagger$	& $	0.656	\pm	0.021	$ & $	25.74	\pm	1.19	$ & $	^\star					$ & $	^\star	$ & yes \\
					&	09.03.2015	& $	0.644	\pm	0.007	$ & $	28.40	\pm	1.86	$ & $	-	$ & $	4.44	\pm	0.32			$ & \\
		HAT-P-28	&	21.10.2014	& $	0.972	\pm	0.019	$ & $	212.34	\pm	2.05	$ & $	6.2:	$ & $	4.09	\pm	0.27	$ & yes \\
		HAT-P-29	&	06.03.2015	& $	3.276	\pm	0.104	$ & $	160.71	\pm	1.36	$ & $	7.93	\pm	0.25	$ & $	6.73	\pm	0.35	$ & yes \\
					&	21.10.2014	& $	3.285	\pm	0.050	$ & $	161.64	\pm	2.36	$ & $	6.31	\pm	0.42	$ & $	6.11	\pm	0.58	$ & \\
		HAT-P-30	&	09.03.2015	& $	3.842	\pm	0.007	$ & $	4.25	\pm	0.14	$ & $	4.50		\pm	0.06	$ & $	4.03	\pm	0.06	$ & \\
		HAT-P-32	&	09.03.2015	& $	2.930	\pm	0.009	$ & $	110.84	\pm	0.43	$ & $	-					$ & $	5.43	\pm	0.16	$ & \\
		HAT-P-35	&	09.03.2015	& $	0.933	\pm	0.010	$ & $	139.75	\pm	1.23	$ & $	5.09	\pm	0.24	$ & $	^{\star, \circ}					$ & yes \\
		HAT-P-41	&	21.10.2014	& $	3.640	\pm	0.011	$ & $	184.00	\pm	0.15	$ & $	3.72		\pm	0.13	$ & $	3.61	\pm	0.17	$ & \\
		HAT-P-54	&	21.10.2014	& $	4.531	\pm	0.062	$ & $	135.95	\pm	1.96	$ & $	5.68	\pm	0.53	$ & $	5.69	\pm	0.59	$ & yes \\
					&	06.03.2015	& $	4.593	\pm	0.010	$ & $	135.82	\pm	0.27	$ & $	^\star					$ & $	5.61	\pm	0.26	$ & \\
		KELT-2		&	21.10.2014	& $	2.402	\pm	0.008	$ & $	332.85	\pm	0.15	$ & $	3.19	\pm	0.09	$ & $	3.13	\pm	0.09	$ & \\
					&	06.03.2015	& $	2.396	\pm	0.007	$ & $	332.81	\pm	0.14	$ & $	2.82	\pm	0.15	$ & $	3.02	\pm	0.15	$ & \\
		KELT-3		&	06.03.2015	& $	3.762	\pm	0.009	$ & $	42.05	\pm	0.23	$ & $	3.93	\pm	0.15	$ & $	3.60		\pm	0.15	$ & \\
		Kepler-13	&	21.10.2014	& $	1.159	\pm	0.003	$ & $	280.02	\pm	0.22	$ & $	0.24	\pm	0.02	$ & $	0.26	\pm	0.03	$ & \\
		Kepler-89	&	21.10.2014	& $	7.316	\pm	0.028	$ & $	108.59	\pm	0.11	$ & $	3.66	\pm	0.18	$ & $	3.37	\pm	0.23	$ & yes \\
		KIC10905746	&	21.10.2014	& $	4.053	\pm	0.007	$ & $	98.61	\pm	0.12	$ & $	2.18	\pm	0.16	$ & $	1.91	\pm	0.16	$ & \\
		LHS-6343	&	21.10.2014	& $	0.723	\pm	0.006	$ & $	119.86	\pm	1.30	$ & $	2.29		\pm	0.24	$ & $	1.66	\pm	0.13	$ & \\
		TrES-4		&	07.03.2015	& $	1.583	\pm	0.019	$ & $	0.69	\pm	0.31	$ & $	4.06	\pm	0.07	$ & $	4.04	\pm	0.09	$ & \\
		WASP-11		&	21.10.2014	& $	0.374	\pm	0.013	$ & $	219.75	\pm	0.79	$ & $	2.94	\pm	0.50	$ & $	2.93	\pm	0.45		$ & \\
					&	06.03.2015	& $	0.383	\pm	0.033	$ & $	218.33	\pm	1.41	$ & $	3.18	\pm	0.26	$ & $	2.91	\pm	0.20		$ & \\
		WASP-12		&	06.03.2015	& $	1.078	\pm	0.008	$ & $	250.08	\pm	0.55	$ & $	4.13	\pm	0.10	$ & $	3.68	\pm	0.08		$ & \\
		WASP-14		&	06.03.2015	& $	1.425	\pm	0.024	$ & $	102.39	\pm	0.40	$ & $	7.14		\pm	0.22	$ & $	5.95	\pm	0.10		$ & \\
		WASP-33		&	21.10.2014	& $	1.920	\pm	0.012	$ & $	275.87	\pm	0.71	$ & $	9.7:	$ & $	7.23	\pm	0.22		$ & \\
		WASP-36		&	09.03.2015	& $	4.845	\pm	0.017	$ & $	67.23	\pm	0.95	$ & $	8.5:	$ & $	6.45	\pm	0.59	$ & \\
		WASP-56		&	06.03.2015	& $	3.424	\pm	0.009	$ & $	113.35	\pm	0.18	$ & $	6.85	\pm	0.24	$ & $	5.95	\pm	0.22	$ & yes \\
		WASP-70		&	21.10.2014	& $	3.195	\pm	0.029	$ & $	167.83	\pm	0.19	$ & $	2.62	\pm	0.18	$ & $	2.49	\pm	0.20		$ & \\
		WASP-76		&	21.10.2014	& $	0.425	\pm	0.012	$ & $	216.90	\pm	2.93	$ & $	2.51	\pm	0.25	$ & $	2.85	\pm	0.33	$ & yes \\
		WASP-77		&	21.10.2014	& $	3.282	\pm	0.007	$ & $	154.02	\pm	0.12	$ & $	1.80		\pm	0.06	$ & $	1.63	\pm	0.06	$ & \\
		WASP-85		&	09.03.2015	& $	1.470	\pm	0.003	$ & $	100.09	\pm	0.19	$ & $	0.89	\pm	0.01	$ & $	0.85	\pm	0.01	$ & \\
		WASP-103	&	07.03.2015	& $	0.242	\pm	0.016	$ & $	132.66	\pm	2.74	$ & $	3.11	\pm	0.46	$ & $	2.59	\pm	0.35	$ & yes \\
		XO-3		&	06.03.2015	& $	6.078	\pm	0.081	$ & $	297.21	\pm	0.13	$ & $	8.13	\pm	0.28	$ & $	7.93	\pm	0.25	$ & \\
		\hline
  	\end{tabular}
	\begin{quote}
		-: no observation in this band \\
		$^\star$: companion candidate was too weak for flux measurement \\
		$^\dagger$: The source was first not identified by us \citep{Woellert15}, but after the new observation with better contrast it could be located. \\
		$^\circ$: The exposure time in $z^{\prime}$ was five times smaller than the one in $i^{\prime}$. 
	\end{quote}
    \end{center}
\end{table*}

\section{Summary}

In our ongoing Lucky Imaging search for stellar sources close to transiting exoplanet host stars we identified $5$ new, close sources within $1 \arcsec$ to HAT-P-27, HAT-P-28, HAT-P-35, WASP-76, and WASP-103 which have been overlooked so far. The planetary and stellar parameters and thermal radiation profile of the transiting planets of these sources may have to be corrected according to the spectral type of the companion candidate star which remains to be determined. Also the detected companion candidates to HAT-P-29 and WASP-56 which are located at $3.3 \arcsec$ and $3.4 \arcsec$ respectively to the TEP host could have this influence as the photometric aperture used for the transit observations, e.g. with SPITZER, are about that size. The sources that are outside of $4 \arcsec$ to HAT-P-15, HAT-P-54, and Kepler-89 do not influence the planetary and stellar property derivation from transit light curve, but can be of interest if they happen to be bound to the TEP host. This needs to be investigated with astrometric observations over the upcoming years. In this work, we also give the astrometric positions and $i^{\prime}$ and $z^{\prime}$ flux ratios of $23$ already known companion candidates.

\begin{acknowledgements}

MW acknowledges support by the International Max Planck Research School for Astronomy \& Cosmic Physics in Heidelberg (IMPRS-HD).

\end{acknowledgements}

\bibliographystyle{aa}
\bibliography{lit}

\begin{thebibliography}{59}
\expandafter\ifx\csname natexlab\endcsname\relax\def\natexlab#1{#1}\fi

\bibitem[{{Adams} {et~al.}(2012){Adams}, {Ciardi}, {Dupree}, {Gautier},
  {Kulesa}, \& {McCarthy}}]{Adams12}
{Adams}, E.~R., {Ciardi}, D.~R., {Dupree}, A.~K., {et~al.} 2012, \aj, 144, 42

\bibitem[{{Adams} {et~al.}(2013){Adams}, {Dupree}, {Kulesa}, \&
  {McCarthy}}]{Adams13}
{Adams}, E.~R., {Dupree}, A.~K., {Kulesa}, C., \& {McCarthy}, D. 2013, \aj,
  146, 9

\bibitem[{{Alonso} {et~al.}(2008){Alonso}, {Auvergne}, {Baglin}, {Ollivier},
  {Moutou}, {Rouan}, {Deeg}, {Aigrain}, {Almenara}, {Barbieri}, {Barge},
  {Benz}, {Bord{\'e}}, {Bouchy}, {de La Reza}, {Deleuil}, {Dvorak}, {Erikson},
  {Fridlund}, {Gillon}, {Gondoin}, {Guillot}, {Hatzes}, {H{\'e}brard},
  {Kabath}, {Jorda}, {Lammer}, {L{\'e}ger}, {Llebaria}, {Loeillet}, {Magain},
  {Mayor}, {Mazeh}, {P{\"a}tzold}, {Pepe}, {Pont}, {Queloz}, {Rauer},
  {Shporer}, {Schneider}, {Stecklum}, {Udry}, \& {Wuchterl}}]{Alonso08}
{Alonso}, R., {Auvergne}, M., {Baglin}, A., {et~al.} 2008, \aap, 482, L21

\bibitem[{{Anderson} {et~al.}(2014){Anderson}, {Collier Cameron}, {Delrez},
  {Doyle}, {Faedi}, {Fumel}, {Gillon}, {G{\'o}mez Maqueo Chew}, {Hellier},
  {Jehin}, {Lendl}, {Maxted}, {Pepe}, {Pollacco}, {Queloz}, {S{\'e}gransan},
  {Skillen}, {Smalley}, {Smith}, {Southworth}, {Triaud}, {Turner}, {Udry}, \&
  {West}}]{Anderson14}
{Anderson}, D.~R., {Collier Cameron}, A., {Delrez}, L., {et~al.} 2014, \mnras,
  445, 1114

\bibitem[{{Artymowicz} \& {Lubow}(1994)}]{Artymowicz94}
{Artymowicz}, P. \& {Lubow}, S.~H. 1994, \apj, 421, 651

\bibitem[{{Baglin} {et~al.}(2006){Baglin}, {Auvergne}, {Barge}, {Deleuil},
  {Catala}, {Michel}, {Weiss}, \& {COROT Team}}]{Baglin06}
{Baglin}, A., {Auvergne}, M., {Barge}, P., {et~al.} 2006, in ESA Special
  Publication, Vol. 1306, ESA Special Publication, ed. M.~{Fridlund},
  A.~{Baglin}, J.~{Lochard}, \& L.~{Conroy}, 33

\bibitem[{{Bakos} {et~al.}(2004){Bakos}, {Noyes}, {Kov{\'a}cs}, {Stanek},
  {Sasselov}, \& {Domsa}}]{Bakos04}
{Bakos}, G., {Noyes}, R.~W., {Kov{\'a}cs}, G., {et~al.} 2004, \pasp, 116, 266

\bibitem[{{Bakos} {et~al.}(2011){Bakos}, {Hartman}, {Torres}, {Latham},
  {Kov{\'a}cs}, {Noyes}, {Fischer}, {Johnson}, {Marcy}, {Howard}, {Kipping},
  {Esquerdo}, {Shporer}, {B{\'e}ky}, {Buchhave}, {Perumpilly}, {Everett},
  {Sasselov}, {Stefanik}, {L{\'a}z{\'a}r}, {Papp}, \& {S{\'a}ri}}]{Bakos11}
{Bakos}, G.~{\'A}., {Hartman}, J., {Torres}, G., {et~al.} 2011, \apj, 742, 116

\bibitem[{{Batalha} {et~al.}(2013){Batalha}, {Rowe}, {Bryson}, {Barclay},
  {Burke}, {Caldwell}, {Christiansen}, {Mullally}, {Thompson}, {Brown},
  {Dupree}, {Fabrycky}, {Ford}, {Fortney}, {Gilliland}, {Isaacson}, {Latham},
  {Marcy}, {Quinn}, {Ragozzine}, {Shporer}, {Borucki}, {Ciardi}, {Gautier},
  {Haas}, {Jenkins}, {Koch}, {Lissauer}, {Rapin}, {Basri}, {Boss}, {Buchhave},
  {Carter}, {Charbonneau}, {Christensen-Dalsgaard}, {Clarke}, {Cochran},
  {Demory}, {Desert}, {Devore}, {Doyle}, {Esquerdo}, {Everett}, {Fressin},
  {Geary}, {Girouard}, {Gould}, {Hall}, {Holman}, {Howard}, {Howell},
  {Ibrahim}, {Kinemuchi}, {Kjeldsen}, {Klaus}, {Li}, {Lucas}, {Meibom},
  {Morris}, {Pr{\v s}a}, {Quintana}, {Sanderfer}, {Sasselov}, {Seader},
  {Smith}, {Steffen}, {Still}, {Stumpe}, {Tarter}, {Tenenbaum}, {Torres},
  {Twicken}, {Uddin}, {Van Cleve}, {Walkowicz}, \& {Welsh}}]{Batalha13}
{Batalha}, N.~M., {Rowe}, J.~F., {Bryson}, S.~T., {et~al.} 2013, \apjs, 204, 24

\bibitem[{{Batygin}(2012)}]{Batygin12}
{Batygin}, K. 2012, \nat, 491, 418

\bibitem[{{Beatty} {et~al.}(2012){Beatty}, {Pepper}, {Siverd}, {Eastman},
  {Bieryla}, {Latham}, {Buchhave}, {Jensen}, {Manner}, {Stassun}, {Gaudi},
  {Berlind}, {Calkins}, {Collins}, {DePoy}, {Esquerdo}, {Fulton}, {F{\H
  u}r{\'e}sz}, {Geary}, {Gould}, {Hebb}, {Kielkopf}, {Marshall}, {Pogge},
  {Stanek}, {Stefanik}, {Street}, {Szentgyorgyi}, {Trueblood}, {Trueblood}, \&
  {Stutz}}]{Beatty12}
{Beatty}, T.~G., {Pepper}, J., {Siverd}, R.~J., {et~al.} 2012, \apjl, 756, L39

\bibitem[{{Bechter} {et~al.}(2014){Bechter}, {Crepp}, {Ngo}, {Knutson},
  {Batygin}, {Hinkley}, {Muirhead}, {Johnson}, {Howard}, {Montet}, {Matthews},
  \& {Morton}}]{Bechter14}
{Bechter}, E.~B., {Crepp}, J.~R., {Ngo}, H., {et~al.} 2014, \apj, 788, 2

\bibitem[{{Bergfors} {et~al.}(2013){Bergfors}, {Brandner}, {Daemgen}, {Biller},
  {Hippler}, {Janson}, {Kudryavtseva}, {Gei{\ss}ler}, {Henning}, \&
  {K{\"o}hler}}]{Bergfors13}
{Bergfors}, C., {Brandner}, W., {Daemgen}, S., {et~al.} 2013, \mnras, 428, 182

\bibitem[{{Bieryla} {et~al.}(2015){Bieryla}, {Collins}, {Beatty}, {Eastman},
  {Siverd}, {Pepper}, {Gaudi}, {Stassun}, {Canas}, {Latham}, {Buchhave},
  {Sanchis-Ojeda}, {Winn}, {Jensen}, {Kielkopf}, {McLeod}, {Gregorio}, {Colon},
  {Street}, {Ross}, {Penny}, {Mellon}, {Oberst}, {Fulton}, {Wang}, {Berlind},
  {Calkins}, {Esquerdo}, {DePoy}, {Gould}, {Marshall}, {Pogge}, {Trueblood}, \&
  {Trueblood}}]{Bieryla15}
{Bieryla}, A., {Collins}, K., {Beatty}, T.~G., {et~al.} 2015, ArXiv e-prints

\bibitem[{{Borucki} {et~al.}(2010){Borucki}, {Koch}, {Basri}, {Batalha},
  {Brown}, {Caldwell}, {Caldwell}, {Christensen-Dalsgaard}, {Cochran},
  {DeVore}, {Dunham}, {Dupree}, {Gautier}, {Geary}, {Gilliland}, {Gould},
  {Howell}, {Jenkins}, {Kondo}, {Latham}, {Marcy}, {Meibom}, {Kjeldsen},
  {Lissauer}, {Monet}, {Morrison}, {Sasselov}, {Tarter}, {Boss}, {Brownlee},
  {Owen}, {Buzasi}, {Charbonneau}, {Doyle}, {Fortney}, {Ford}, {Holman},
  {Seager}, {Steffen}, {Welsh}, {Rowe}, {Anderson}, {Buchhave}, {Ciardi},
  {Walkowicz}, {Sherry}, {Horch}, {Isaacson}, {Everett}, {Fischer}, {Torres},
  {Johnson}, {Endl}, {MacQueen}, {Bryson}, {Dotson}, {Haas}, {Kolodziejczak},
  {Van Cleve}, {Chandrasekaran}, {Twicken}, {Quintana}, {Clarke}, {Allen},
  {Li}, {Wu}, {Tenenbaum}, {Verner}, {Bruhweiler}, {Barnes}, \&
  {Prsa}}]{Borucki10}
{Borucki}, W.~J., {Koch}, D., {Basri}, G., {et~al.} 2010, Science, 327, 977

\bibitem[{{Brown} {et~al.}(2014){Brown}, {Anderson}, {Armstrong}, {Bouchy},
  {Collier Cameron}, {Delrez}, {Doyle}, {Gillon}, {Gomez Maqueo Chew}, {Hebb},
  {Hebrard}, {Hellier}, {Jehin}, {Lendl}, {Maxted}, {McCormac},
  {Neveu-VanMalle}, {Pollacco}, {Queloz}, {Segransan}, {Smalley}, {Turner},
  {Triaud}, \& {Udry}}]{Brown14}
{Brown}, D.~J.~A., {Anderson}, D.~R., {Armstrong}, D.~J., {et~al.} 2014, ArXiv
  e-prints

\bibitem[{{Burke} {et~al.}(2014){Burke}, {Bryson}, {Mullally}, {Rowe},
  {Christiansen}, {Thompson}, {Coughlin}, {Haas}, {Batalha}, {Caldwell},
  {Jenkins}, {Still}, {Barclay}, {Borucki}, {Chaplin}, {Ciardi}, {Clarke},
  {Cochran}, {Demory}, {Esquerdo}, {Gautier}, {Gilliland}, {Girouard}, {Havel},
  {Henze}, {Howell}, {Huber}, {Latham}, {Li}, {Morehead}, {Morton}, {Pepper},
  {Quintana}, {Ragozzine}, {Seader}, {Shah}, {Shporer}, {Tenenbaum}, {Twicken},
  \& {Wolfgang}}]{Burke14}
{Burke}, C.~J., {Bryson}, S.~T., {Mullally}, F., {et~al.} 2014, \apjs, 210, 19

\bibitem[{{Collins} {et~al.}(2014){Collins}, {Eastman}, {Beatty}, {Siverd},
  {Gaudi}, {Pepper}, {Kielkopf}, {Johnson}, {Howard}, {Fischer}, {Manner},
  {Bieryla}, {Latham}, {Fulton}, {Gregorio}, {Buchhave}, {Jensen}, {Stassun},
  {Penev}, {Crepp}, {Hinkley}, {Street}, {Cargile}, {Mack}, {Oberst}, {Avril},
  {Mellon}, {McLeod}, {Penny}, {Stefanik}, {Berlind}, {Calkins}, {Mao},
  {Richert}, {DePoy}, {Esquerdo}, {Gould}, {Marshall}, {Oelkers}, {Pogge},
  {Trueblood}, \& {Trueblood}}]{Collins14}
{Collins}, K.~A., {Eastman}, J.~D., {Beatty}, T.~G., {et~al.} 2014, \aj, 147,
  39

\bibitem[{{Col{\'o}n} {et~al.}(2012){Col{\'o}n}, {Ford}, \&
  {Morehead}}]{Colon12}
{Col{\'o}n}, K.~D., {Ford}, E.~B., \& {Morehead}, R.~C. 2012, \mnras, 426, 342

\bibitem[{{Crossfield} {et~al.}(2012){Crossfield}, {Barman}, {Hansen},
  {Tanaka}, \& {Kodama}}]{Crossfield12}
{Crossfield}, I.~J.~M., {Barman}, T., {Hansen}, B.~M.~S., {Tanaka}, I., \&
  {Kodama}, T. 2012, \apj, 760, 140

\bibitem[{{Crossfield} {et~al.}(2015){Crossfield}, {Petigura}, {Schlieder},
  {Howard}, {Fulton}, {Aller}, {Ciardi}, {L{\'e}pine}, {Barclay}, {de Pater},
  {de Kleer}, {Quintana}, {Christiansen}, {Schlafly}, {Kaltenegger}, {Crepp},
  {Henning}, {Obermeier}, {Deacon}, {Weiss}, {Isaacson}, {Hansen}, {Liu},
  {Greene}, {Howell}, {Barman}, \& {Mordasini}}]{Crossfield15}
{Crossfield}, I.~J.~M., {Petigura}, E., {Schlieder}, J.~E., {et~al.} 2015,
  \apj, 804, 10

\bibitem[{{Daemgen} {et~al.}(2009){Daemgen}, {Hormuth}, {Brandner}, {Bergfors},
  {Janson}, {Hippler}, \& {Henning}}]{Daemgen09}
{Daemgen}, S., {Hormuth}, F., {Brandner}, W., {et~al.} 2009, \aap, 498, 567

\bibitem[{{Deleuil} {et~al.}(2008){Deleuil}, {Deeg}, {Alonso}, {Bouchy},
  {Rouan}, {Auvergne}, {Baglin}, {Aigrain}, {Almenara}, {Barbieri}, {Barge},
  {Bruntt}, {Bord{\'e}}, {Collier Cameron}, {Csizmadia}, {de La Reza},
  {Dvorak}, {Erikson}, {Fridlund}, {Gandolfi}, {Gillon}, {Guenther}, {Guillot},
  {Hatzes}, {H{\'e}brard}, {Jorda}, {Lammer}, {L{\'e}ger}, {Llebaria},
  {Loeillet}, {Mayor}, {Mazeh}, {Moutou}, {Ollivier}, {P{\"a}tzold}, {Pont},
  {Queloz}, {Rauer}, {Schneider}, {Shporer}, {Wuchterl}, \&
  {Zucker}}]{Deleuil08}
{Deleuil}, M., {Deeg}, H.~J., {Alonso}, R., {et~al.} 2008, \aap, 491, 889

\bibitem[{{D{\'e}sert} {et~al.}(2015){D{\'e}sert}, {Charbonneau}, {Torres},
  {Fressin}, {Ballard}, {Bryson}, {Knutson}, {Batalha}, {Borucki}, {Brown},
  {Deming}, {Ford}, {Fortney}, {Gilliland}, {Latham}, \& {Seager}}]{Desert15}
{D{\'e}sert}, J.-M., {Charbonneau}, D., {Torres}, G., {et~al.} 2015, \apj, 804,
  59

\bibitem[{{Dressing} {et~al.}(2014){Dressing}, {Adams}, {Dupree}, {Kulesa}, \&
  {McCarthy}}]{Dressing14}
{Dressing}, C.~D., {Adams}, E.~R., {Dupree}, A.~K., {Kulesa}, C., \&
  {McCarthy}, D. 2014, \aj, 148, 78

\bibitem[{{Everett} {et~al.}(2015){Everett}, {Barclay}, {Ciardi}, {Horch},
  {Howell}, {Crepp}, \& {Silva}}]{Everett15}
{Everett}, M.~E., {Barclay}, T., {Ciardi}, D.~R., {et~al.} 2015, \aj, 149, 55

\bibitem[{{Fabrycky} \& {Tremaine}(2007)}]{Fabrycky07}
{Fabrycky}, D. \& {Tremaine}, S. 2007, \apj, 669, 1298

\bibitem[{{Faedi} {et~al.}(2013){Faedi}, {Staley}, {G{\'o}mez Maqueo Chew},
  {Pollacco}, {Dhital}, {Barros}, {Skillen}, {Hebb}, {Mackay}, \&
  {Watson}}]{Faedi13}
{Faedi}, F., {Staley}, T., {G{\'o}mez Maqueo Chew}, Y., {et~al.} 2013, \mnras,
  433, 2097

\bibitem[{{Fischer} {et~al.}(2012){Fischer}, {Schwamb}, {Schawinski},
  {Lintott}, {Brewer}, {Giguere}, {Lynn}, {Parrish}, {Sartori}, {Simpson},
  {Smith}, {Spronck}, {Batalha}, {Rowe}, {Jenkins}, {Bryson}, {Prsa},
  {Tenenbaum}, {Crepp}, {Morton}, {Howard}, {Beleu}, {Kaplan}, {Vannispen},
  {Sharzer}, {Defouw}, {Hajduk}, {Neal}, {Nemec}, {Schuepbach}, \&
  {Zimmermann}}]{Fischer12}
{Fischer}, D.~A., {Schwamb}, M.~E., {Schawinski}, K., {et~al.} 2012, \mnras,
  419, 2900

\bibitem[{{Gandolfi} {et~al.}(2010){Gandolfi}, {H{\'e}brard}, {Alonso},
  {Deleuil}, {Guenther}, {Fridlund}, {Endl}, {Eigm{\"u}ller}, {Csizmadia},
  {Havel}, {Aigrain}, {Auvergne}, {Baglin}, {Barge}, {Bonomo}, {Bord{\'e}},
  {Bouchy}, {Bruntt}, {Cabrera}, {Carpano}, {Carone}, {Cochran}, {Deeg},
  {Dvorak}, {Eisl{\"o}ffel}, {Erikson}, {Ferraz-Mello}, {Gazzano}, {Gibson},
  {Gillon}, {Gondoin}, {Guillot}, {Hartmann}, {Hatzes}, {Jorda}, {Kabath},
  {L{\'e}ger}, {Llebaria}, {Lammer}, {MacQueen}, {Mayor}, {Mazeh}, {Moutou},
  {Ollivier}, {P{\"a}tzold}, {Pepe}, {Queloz}, {Rauer}, {Rouan}, {Samuel},
  {Schneider}, {Stecklum}, {Tingley}, {Udry}, \& {Wuchterl}}]{Gandolfi10}
{Gandolfi}, D., {H{\'e}brard}, G., {Alonso}, R., {et~al.} 2010, \aap, 524, A55

\bibitem[{{Guenther} {et~al.}(2013){Guenther}, {Fridlund}, {Alonso}, {Carpano},
  {Deeg}, {Deleuil}, {Dreizler}, {Endl}, {Gandolfi}, {Gillon}, {Guillot},
  {Jehin}, {L{\'e}ger}, {Moutou}, {Nortmann}, {Rouan}, {Samuel}, {Schneider},
  \& {Tingley}}]{Guenther13}
{Guenther}, E.~W., {Fridlund}, M., {Alonso}, R., {et~al.} 2013, \aap, 556, A75

\bibitem[{{Hartman} {et~al.}(2012){Hartman}, {Bakos}, {B{\'e}ky}, {Torres},
  {Latham}, {Csubry}, {Penev}, {Shporer}, {Fulton}, {Buchhave}, {Johnson},
  {Howard}, {Marcy}, {Fischer}, {Kov{\'a}cs}, {Noyes}, {Esquerdo}, {Everett},
  {Szklen{\'a}r}, {Quinn}, {Bieryla}, {Knox}, {Hinz}, {Sasselov}, {F{\H
  u}r{\'e}sz}, {Stefanik}, {L{\'a}z{\'a}r}, {Papp}, \& {S{\'a}ri}}]{Hartman12}
{Hartman}, J.~D., {Bakos}, G.~{\'A}., {B{\'e}ky}, B., {et~al.} 2012, \aj, 144,
  139

\bibitem[{{Horch} {et~al.}(2014){Horch}, {Howell}, {Everett}, \&
  {Ciardi}}]{Horch14}
{Horch}, E.~P., {Howell}, S.~B., {Everett}, M.~E., \& {Ciardi}, D.~R. 2014,
  \apj, 795, 60

\bibitem[{{Hormuth} {et~al.}(2008){Hormuth}, {Brandner}, {Hippler}, \&
  {Henning}}]{Hormuth08}
{Hormuth}, F., {Brandner}, W., {Hippler}, S., \& {Henning}, T. 2008, Journal of
  Physics Conference Series, 131, 012051

\bibitem[{{Howell} {et~al.}(2011){Howell}, {Everett}, {Sherry}, {Horch}, \&
  {Ciardi}}]{Howell11}
{Howell}, S.~B., {Everett}, M.~E., {Sherry}, W., {Horch}, E., \& {Ciardi},
  D.~R. 2011, \aj, 142, 19

\bibitem[{{Johnson} {et~al.}(2011){Johnson}, {Apps}, {Gazak}, {Crepp},
  {Crossfield}, {Howard}, {Marcy}, {Morton}, {Chubak}, \&
  {Isaacson}}]{Johnson11}
{Johnson}, J.~A., {Apps}, K., {Gazak}, J.~Z., {et~al.} 2011, \apj, 730, 79

\bibitem[{{Kane} {et~al.}(2014){Kane}, {Howell}, {Horch}, {Feng}, {Hinkel},
  {Ciardi}, {Everett}, {Howard}, \& {Wright}}]{Kane14}
{Kane}, S.~R., {Howell}, S.~B., {Horch}, E.~P., {et~al.} 2014, \apj, 785, 93

\bibitem[{{Law} {et~al.}(2014){Law}, {Morton}, {Baranec}, {Riddle},
  {Ravichandran}, {Ziegler}, {Johnson}, {Tendulkar}, {Bui}, {Burse}, {Das},
  {Dekany}, {Kulkarni}, {Punnadi}, \& {Ramaprakash}}]{Law14}
{Law}, N.~M., {Morton}, T., {Baranec}, C., {et~al.} 2014, \apj, 791, 35

\bibitem[{{Lillo-Box} {et~al.}(2012){Lillo-Box}, {Barrado}, \&
  {Bouy}}]{Lillo-Box12}
{Lillo-Box}, J., {Barrado}, D., \& {Bouy}, H. 2012, \aap, 546, A10

\bibitem[{{Lillo-Box} {et~al.}(2014){Lillo-Box}, {Barrado}, \&
  {Bouy}}]{Lillo-Box14}
{Lillo-Box}, J., {Barrado}, D., \& {Bouy}, H. 2014, \aap, 566, A103

\bibitem[{{Maxted} {et~al.}(2013){Maxted}, {Anderson}, {Collier Cameron},
  {Doyle}, {Fumel}, {Gillon}, {Hellier}, {Jehin}, {Lendl}, {Pepe}, {Pollacco},
  {Queloz}, {S{\'e}gransan}, {Smalley}, {Southworth}, {Smith}, {Triaud},
  {Udry}, \& {West}}]{Maxted13}
{Maxted}, P.~F.~L., {Anderson}, D.~R., {Collier Cameron}, A., {et~al.} 2013,
  \pasp, 125, 48

\bibitem[{{Mayer} {et~al.}(2005){Mayer}, {Wadsley}, {Quinn}, \&
  {Stadel}}]{Mayer05}
{Mayer}, L., {Wadsley}, J., {Quinn}, T., \& {Stadel}, J. 2005, \mnras, 363, 641

\bibitem[{{Montet} {et~al.}(2015){Montet}, {Johnson}, {Muirhead}, {Villar},
  {Vassallo}, {Baranec}, {Law}, {Riddle}, {Marcy}, {Howard}, \&
  {Isaacson}}]{Montet15}
{Montet}, B.~T., {Johnson}, J.~A., {Muirhead}, P.~S., {et~al.} 2015, \apj, 800,
  134

\bibitem[{{Moya} {et~al.}(2011){Moya}, {Bouy}, {Marchis}, {Vicente}, \&
  {Barrado}}]{Moya11}
{Moya}, A., {Bouy}, H., {Marchis}, F., {Vicente}, B., \& {Barrado}, D. 2011,
  \aap, 535, A110

\bibitem[{{Naoz} {et~al.}(2011){Naoz}, {Farr}, {Lithwick}, {Rasio}, \&
  {Teyssandier}}]{Naoz11}
{Naoz}, S., {Farr}, W.~M., {Lithwick}, Y., {Rasio}, F.~A., \& {Teyssandier}, J.
  2011, \nat, 473, 187

\bibitem[{{Narita} {et~al.}(2012){Narita}, {Takahashi}, {Kuzuhara}, {Hirano},
  {Suenaga}, {Kandori}, {Kudo}, {Sato}, {Suzuki}, {Ida}, {Nagasawa}, {Abe},
  {Brandner}, {Brandt}, {Carson}, {Egner}, {Feldt}, {Goto}, {Grady}, {Guyon},
  {Hashimoto}, {Hayano}, {Hayashi}, {Hayashi}, {Henning}, {Hodapp}, {Ishii},
  {Iye}, {Janson}, {Knapp}, {Kusakabe}, {Kwon}, {Matsuo}, {Mayama}, {McElwain},
  {Miyama}, {Morino}, {Moro-Martin}, {Nishimura}, {Pyo}, {Serabyn}, {Suto},
  {Takami}, {Takato}, {Terada}, {Thalmann}, {Tomono}, {Turner}, {Watanabe},
  {Wisniewski}, {Yamada}, {Takami}, {Usuda}, \& {Tamura}}]{Narita12}
{Narita}, N., {Takahashi}, Y.~H., {Kuzuhara}, M., {et~al.} 2012, \pasj, 64, L7

\bibitem[{{Ngo} {et~al.}(2015){Ngo}, {Knutson}, {Hinkley}, {Crepp}, {Bechter},
  {Batygin}, {Howard}, {Johnson}, {Morton}, \& {Muirhead}}]{NGO15}
{Ngo}, H., {Knutson}, H.~A., {Hinkley}, S., {et~al.} 2015, \apj, 800, 138

\bibitem[{{Pepper} {et~al.}(2013){Pepper}, {Siverd}, {Beatty}, {Gaudi},
  {Stassun}, {Eastman}, {Collins}, {Latham}, {Bieryla}, {Buchhave}, {Jensen},
  {Manner}, {Penev}, {Crepp}, {Cargile}, {Dhital}, {Calkins}, {Esquerdo},
  {Berlind}, {Fulton}, {Street}, {Ma}, {Ge}, {Wang}, {Mao}, {Richert}, {Gould},
  {DePoy}, {Kielkopf}, {Marshall}, {Pogge}, {Stefanik}, {Trueblood}, \&
  {Trueblood}}]{Pepper13}
{Pepper}, J., {Siverd}, R.~J., {Beatty}, T.~G., {et~al.} 2013, \apj, 773, 64

\bibitem[{{Pollacco} {et~al.}(2006){Pollacco}, {Skillen}, {Collier Cameron},
  {Christian}, {Hellier}, {Irwin}, {Lister}, {Street}, {West}, {Anderson},
  {Clarkson}, {Deeg}, {Enoch}, {Evans}, {Fitzsimmons}, {Haswell}, {Hodgkin},
  {Horne}, {Kane}, {Keenan}, {Maxted}, {Norton}, {Osborne}, {Parley}, {Ryans},
  {Smalley}, {Wheatley}, \& {Wilson}}]{Pollacco06}
{Pollacco}, D.~L., {Skillen}, I., {Collier Cameron}, A., {et~al.} 2006, \pasp,
  118, 1407

\bibitem[{{Roell} {et~al.}(2012){Roell}, {Neuh{\"a}user}, {Seifahrt}, \&
  {Mugrauer}}]{Roell12}
{Roell}, T., {Neuh{\"a}user}, R., {Seifahrt}, A., \& {Mugrauer}, M. 2012, \aap,
  542, A92

\bibitem[{{Sanchis-Ojeda} {et~al.}(2013){Sanchis-Ojeda}, {Winn}, {Marcy},
  {Howard}, {Isaacson}, {Johnson}, {Torres}, {Albrecht}, {Campante}, {Chaplin},
  {Davies}, {Lund}, {Carter}, {Dawson}, {Buchhave}, {Everett}, {Fischer},
  {Geary}, {Gilliland}, {Horch}, {Howell}, \& {Latham}}]{Sanchis-Ojeda13}
{Sanchis-Ojeda}, R., {Winn}, J.~N., {Marcy}, G.~W., {et~al.} 2013, \apj, 775,
  54

\bibitem[{{Santerne} {et~al.}(2012){Santerne}, {Moutou}, {Barros}, {Damiani},
  {D{\'{\i}}az}, {Almenara}, {Bonomo}, {Bouchy}, {Deleuil}, \&
  {H{\'e}brard}}]{Santerne12}
{Santerne}, A., {Moutou}, C., {Barros}, S.~C.~C., {et~al.} 2012, \aap, 544, L12

\bibitem[{{Siverd} {et~al.}(2012){Siverd}, {Beatty}, {Pepper}, {Eastman},
  {Collins}, {Bieryla}, {Latham}, {Buchhave}, {Jensen}, {Crepp}, {Street},
  {Stassun}, {Gaudi}, {Berlind}, {Calkins}, {DePoy}, {Esquerdo}, {Fulton},
  {F{\H u}r{\'e}sz}, {Geary}, {Gould}, {Hebb}, {Kielkopf}, {Marshall}, {Pogge},
  {Stanek}, {Stefanik}, {Szentgyorgyi}, {Trueblood}, {Trueblood}, {Stutz}, \&
  {van Saders}}]{Siverd12}
{Siverd}, R.~J., {Beatty}, T.~G., {Pepper}, J., {et~al.} 2012, \apj, 761, 123

\bibitem[{{Smith} {et~al.}(2012){Smith}, {Anderson}, {Collier Cameron},
  {Gillon}, {Hellier}, {Lendl}, {Maxted}, {Queloz}, {Smalley}, {Triaud},
  {West}, {Barros}, {Jehin}, {Pepe}, {Pollacco}, {Segransan}, {Southworth},
  {Street}, \& {Udry}}]{Smith12}
{Smith}, A.~M.~S., {Anderson}, D.~R., {Collier Cameron}, A., {et~al.} 2012,
  \aj, 143, 81

\bibitem[{{Wang} {et~al.}(2014){Wang}, {Fischer}, {Xie}, \& {Ciardi}}]{Wang14}
{Wang}, J., {Fischer}, D.~A., {Xie}, J.-W., \& {Ciardi}, D.~R. 2014, \apj, 791,
  111

\bibitem[{{Winn} {et~al.}(2005){Winn}, {Noyes}, {Holman}, {Charbonneau},
  {Ohta}, {Taruya}, {Suto}, {Narita}, {Turner}, {Johnson}, {Marcy}, {Butler},
  \& {Vogt}}]{Winn05}
{Winn}, J.~N., {Noyes}, R.~W., {Holman}, M.~J., {et~al.} 2005, \apj, 631, 1215

\bibitem[{{W{\"o}llert} {et~al.}(2015){W{\"o}llert}, {Brandner}, {Bergfors}, \&
  {Henning}}]{Woellert15}
{W{\"o}llert}, M., {Brandner}, W., {Bergfors}, C., \& {Henning}, T. 2015, \aap,
  575, A23

\bibitem[{{W{\"o}llert} {et~al.}(2014){W{\"o}llert}, {Brandner}, {Reffert},
  {Schlieder}, {Mohler-Fischer}, {K{\"o}hler}, \& {Henning}}]{Woellert14}
{W{\"o}llert}, M., {Brandner}, W., {Reffert}, S., {et~al.} 2014, \aap, 564, A10

\bibitem[{{Wu} \& {Murray}(2003)}]{Wu03}
{Wu}, Y. \& {Murray}, N. 2003, \apj, 589, 605

\end{thebibliography}

\end{document}